\documentclass[journal=jacsat,manuscript=article, ]{achemso}
\usepackage[version=3]{mhchem} 

\author{Thibaut Devillers}
\affiliation[Institut f\"ur Halbleiter-und-Festk\"orperphysik]{Institut f\"ur Halbleiter-und-Festk\"orperphysik, Johannes Kepler University, Altenbergerstr. 69, A-4040 Linz, Austria}
\email{thibaut.devillers@neel.cnrs.fr}

\author{Li Tian}
\affiliation[Institute of Physics, Polish Academy of Science]{Institute of Physics PAS SL 1.4, al. Lotnikow 32/46, 02-668 Warsaw, Poland}

\author{Rajdeep Adhikari}
\affiliation[Institut f\"ur Halbleiter-und-Festk\"orperphysik]{Institut f\"ur Halbleiter-und-Festk\"orperphysik, Johannes Kepler University, Altenbergerstr. 69, A-4040 Linz, Austria}

\author{Giulia Capuzzo}
\affiliation[Institut f\"ur Halbleiter-und-Festk\"orperphysik]{Institut f\"ur Halbleiter-und-Festk\"orperphysik, Johannes Kepler University, Altenbergerstr. 69, A-4040 Linz, Austria}

\author{Alberta Bonanni}
\affiliation[Institut f\"ur Halbleiter-und-Festk\"orperphysik]{Institut f\"ur Halbleiter-und-Festk\"orperphysik, Johannes Kepler University, Altenbergerstr. 69, A-4040 Linz, Austria}
\email{alberta.bonanni@jku.at}

\title{Mn as surfactant for the self-assembling of Al$_x$Ga$_{1-x}$N/GaN layered heterostructures}

\begin{document}

\begin{tocentry}

Some journals require a graphical entry for the Table of Contents.
This should be laid out ``print ready'' so that the sizing of the
text is correct.

Inside the \texttt{tocentry} environment, the font used is Helvetica
8\,pt, as required by \emph{Journal of the American Chemical
Society}.

The surrounding frame is 9\,cm by 3.5\,cm, which is the maximum
permitted for  \emph{Journal of the American Chemical Society}
graphical table of content entries. The box will not resize if the
content is too big: instead it will overflow the edge of the box.

This box and the associated title will always be printed on a
separate page at the end of the document.

\end{tocentry}

\begin{abstract}
The structural analysis of GaN and Al$_x$Ga$_{1-x}$N/GaN heterostructures grown by metalorganic vapor phase epitaxy in the presence of Mn reveals how Mn affects the growth process, and in particular the incorporation of Al, the morphology of the surface, and the plastic relaxation of Al$_x$Ga$_{1-x}$N on GaN. Moreover, the doping with Mn promotes the formation of layered Al$_x$Ga$_{1-x}$N/GaN superlattice-like heterostructures opening wide perspective for controlling the segregation of ternary alloys during the crystal growth and for fostering the self-assembling of functional layered structures.
\end{abstract}

\section{Introduction}
Nitride heterostructures\cite{MorkocBook2008} are the building blocks of many state-of-the-art devices like power transistors\cite{shur1998}, high-electron-mobility transistors\cite{Mishra2002}, blue and white light-emitting diodes (LED)\cite{Gutt2012}, ultra-violet laser diodes\cite{Yoshida2008}, and blue lasers\cite{nakamura2000}. In order to produce a heterostructure whose band structure responds to the actual need, various combinations of nitride compounds with different band gaps and lattice parameters are epitaxially stacked to generate $e.g$ quantum wells\cite{lepkowski2013}, super-lattices, barriers\cite{Manuel2011}, or distributed Bragg reflector\cite{Kruse2011}. In any of these structures, the quality and continuity of the crystal is essential for the performance of the device.  
However, as expected during the epitaxial growth of lattice mismatched semiconductors, active layers of Al$_x$Ga$_{1-x}$N or In$_x$Ga$_{1-x}$N\cite{Ra2013} deposited on GaN as requested by the architecture of most devices will tend to crack in order to relax the elastic strain accumulated by fitting the in-plane lattice parameter of the overlayer to the one of the layer underneath. Growing above the critical thickness is a challenging way to design high-performance devices, with $e.g.$ enhanced internal fields\cite{Songmuang2011,Dong2014} and piezoelectric polarization\cite{Ambacher2000}. In this perspective, the properties of the sample surface may be affected through the use of a surfactant\cite{Copel1989}. The efficiency of this approach was already demonstrated in the case of Te for the growth of InAs on GaAs\cite{Grandjean1992}, and As or Sb for Ge on Si\cite{Osten1992}. Specifically, in the nitride technology, In\cite{Widmann1998,Keller2001,Monroy2002,Nicolay2006}, As\cite{Okumura1998}, Ga\cite{Mula2001} and Sb\cite{Zhang2001,Zhang2002} were reported to have a beneficial effect on the growth of GaN, as well as on the fabrication of AlN/GaN heterostructures.

Here, we report on the role of Mn in the crystal growth of GaN:Mn and Al$_x$Ga$_{1-x}$N:Mn on GaN by metalorganic vapor phase epitaxy (MOVPE). In particular, we investigate how Mn affects both the morphology of the surface, and the bulk crystal structure.

\section{Experimental section}
The studied samples are grown by MOVPE, in an AIXTRON 200RF horizontal reactor, according to a procedure described elsewhere\cite{stefanowicz2010}. A 1\,$\mu$m GaN buffer layer is deposited epitaxially at 1040\,$^\circ$C on sapphire \emph{c}-plane after the growth of a low temperature nucleation layer. The layers studied in this work are then deposited at a temperature of 850\,$^\circ$C on the GaN buffer. This unusually low growth temperature is of technical significance for several reasons. High growth temperatures ($\sim 1000^\circ$C ) on one hand promote  the crystalline quality, but on the other favor the propagation of threading dislocations through the epitaxial films. Lower growth temperature – as employed in this work for the growth of (AlGa)N – hinder the propagation of the dislocations~\cite{Bourret2000}. Furthermore, the difference in thermal expansion coefficient between overlayer and substrates may highly strain the layers during the cooling and  induce cracking of the epitaxial structure. Thermal stress can be reduced significantly by using low growth temperatures. Finally, the integration of nitride technology in devices, may require lower growth temperature, to avoid potentially detrimental diffusion of species during processing. The precursors employed for Ga, N, Al, and Mn are trimethylgallium (TMGa), ammonia (NH$_3$), trimethylaluminium (TMAl), and bis-methylcyclopentadienyl-manganese (MeCp$_2$Mn), respectively. The flow of TMGa and NH$_3$ are fixed at 4~$\mu$mol/min and 7000~$\mu$mol/min respectively. The flow of MeCp$_2$Mn is 1~$\mu$mol/min. The flow of TMAl is varied between 0.4 and 31~$\mu$mol/min. The growth is carried out under H$_2$ atmosphere, at a pressure of 100\,mbar for the Al$_x$Ga$_{1-x}$N (Al$_x$Ga$_{1-x}$N:Mn) layers, and 200\,mbar for the GaN (GaN:Mn). The relevant characteristics of the samples studied here are summarized in table~\ref{table:sample}. 
The thickness of the layers is controlled \emph{in~situ} during the growth process by kinetic ellipsometry and \emph{ex~situ} with spectroscopic ellipsometry, secondary ion mass spectroscopy (SIMS) and x-ray reflectivity. The Al concentration is calculated from the position of the (0002) and $(\overline{1}015)$ diffraction peaks of Al$_x$Ga$_{1-x}$N.

\begin{table}
  \caption{Samples investigated and their relevant parameters.}
  \label{table:sample}
  \begin{tabular}{|c|llllll|}
    \hline
    Sample  	& buffer 	& layer 	& thickness 	& Al concentration 	& growth temperature	& growth pressure \tabularnewline
		   				&					&					& (nm)				& (\%)							& ($^\circ$C)					& (mbar)					\\
    \hline
    \#A  				& GaN 		& GaN							& 500			& 0								& 850 				& 200\\
    \#B 				& GaN 		& GaN:Mn					& 500			& 0								& 850 				& 200\\
    \#C 				& GaN 		& AlGaN						& 15			& 75							& 850 				& 100\\
    \#D					& GaN 		& AlGaN:Mn				& 15			& 75							& 850					& 100\\
    \#E					& GaN 		& AlGaN						& 1000		& 12							& 850 				& 100\\
		\#F					& GaN 		& AlGaN:Mn				& 1000		& 12							& 850					& 100\\
		\hline
  \end{tabular}
\end{table}

Information on the morphology of the surface is obtained from atomic force microscopy (AFM) in tapping mode with a Nanosurf  MobileS and with a VEECO Dimension 3100. X-ray diffraction and reflectivity are performed on a PANalytical's X'Pert PRO Materials Research Diffractometer (MRD) equipped with a hybrid monochromator with a 1/4$^{\circ}$ divergence slit. The diffracted beam is measured with a solid-state PixCel detector used as 256-channels detector with a 11.9\,mm anti-scatter slit. 
Transmission electron microscopy (TEM) in both conventional (CTEM) and scanning mode (STEM) is performed in a FEI Titan Cube
80-300 operating at 300\,keV and in a JEOL 2010F working at 200\,KeV.
Bright/dark-field (BF/DF), high resolution TEM (HRTEM) and high angle annular dark field (HAADF) are employed to analyse the structure of the sample. Chemical mapping is performed with energy filtered TEM (EFTEM), around the Al $\textit{K}$ absorption edge.
Cross-section TEM specimens are prepared by mechanical polishing, dimpling and final ion milling in a Gatan
Precision Ion Polishing System. 

\section{Results and discussion}
\begin{figure}[ht]
  \includegraphics[width=.9\linewidth]{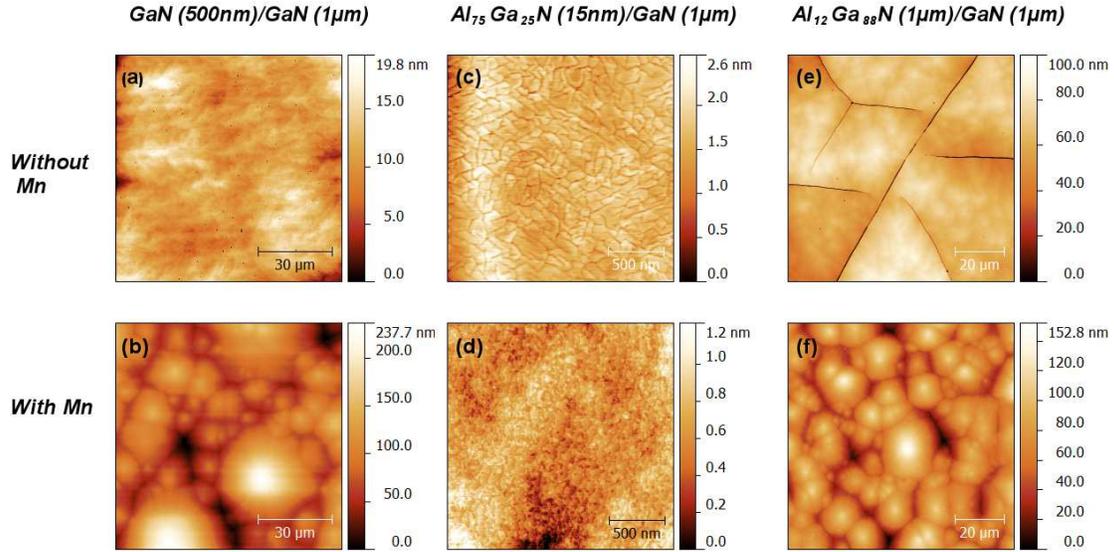}
  \caption{Atomic force micrographs for: (a),(c),(e) GaN and Al$_x$Ga$_{1-x}$N in the absence of Mn; (b),(d),(f) GaN:Mn and Al$_x$Ga$_{1-x}$N:Mn.}
  \label{fig:fig1}
\end{figure}

The effect of Mn on the surface morphology of GaN and Al$_x$Ga$_{1-x}$N is studied by AFM by directly contrasting GaN, Al$_{0.75}$Ga$_{0.25}$N and Al$_{0.12}$Ga$_{0.88}$N grown in the absence of Mn (samples \#A, \#C, and \#E) with samples grown under the same conditions, but deposited in the presence of Mn (samples \#B, \#D, and \#F), as reported in fig.~\ref{fig:fig1}. The compared samples have the same nominal and actual thickness and differ only in the presence of Mn during the growth process.

As evidenced in fig.~\ref{fig:fig1}(a) and (b), the addition of Mn during the growth of GaN at $T_\mathrm{G}$=850$^\circ$C affects the morphology of the surface, inducing the formation of large domes. This effect resembles the one observed by Zhang \emph{et al.}\cite{Zhang2002} during the growth of GaN in the presence of antimony (Sb), where Sb was found to act as a surfactant during the growth, promoting the mobility of Ga atoms on the surface, and thus lateral growth. In our case, the fact that Mn plays the role of a surfactant is coherent with its low probability of incorporation into the crystal. Despite the relatively high flow rate of the MeCp$_2$Mn precursor during the epitaxial process (20\% of the III metal precursor flow), the Mn content in the layers remains of the order of 1\% cations – as established through SIMS measurements – showing that a large part of the Mn is not incorporated in the layer, being either desorbed or accumulated at the surface. The dramatic effect that we do observe on the growth is rather suggesting that an important part of the Mn atoms accumulates at the growth front and influences the dynamics of epitaxy. In addition to the presence of domes, in the samples grown in the presence of Mn, there is no hint of linear (cracks) or punctual (pits) discontinuities of the layers, and one can still distinguish in higher resolution images (not shown) the atomic terrace edges characteristic of a step-flow growth mode. In fig.~\ref{fig:fig1}(c) and (d) the effect of Mn on the growth of 15\,nm thin Al$_{0.75}$Ga$_{0.25}$N layers is evidenced. In the Mn-free layer the expected morphology of Al$_x$Ga$_{1-x}$N close to relaxation\cite{Keller1999} can be appreciated: shallow fractures -- precursors of the cracks observed for relaxed thicker layers -- are detected. Both samples were grown in the same conditions, resulting in the same thickness (as determined from the fitting of x-ray reflectivity) and same Al content (as  obtained from the position of the $(\overline{1}015)$ asymetric peak and from  fitting the x-ray reflectivity). The layer grown in the presence of Mn, even if the thickness and the Al concentration are kept constant, exhibits a surface free of nanofractures, pointing to the fact that Mn delays the relaxation of the lattice like Sb in the case of Ge grown on Si\cite{Osten1992}, or Te in InAs on GaAs\cite{Grandjean1992}. 

For a closer analysis of the role of Mn in the relaxation of Al$_x$Ga$_{1-x}$N on GaN, two layers of Al$_{0.12}$Ga$_{0.88}$N, with and without Mn respectively, and with a thickness of 1\,$\mu$m, $i.e.$ theoretically above the critical thickness\cite{Lee2004} are compared. The Al$_{0.12}$Ga$_{0.88}$N layer grown without Mn and shown in fig.~\ref{fig:fig1}(e) presents surface grooves oriented at either 60$^\circ$ or 120$^\circ$ one with respect to the other. These cracks are characteristic of the heteroepitaxy of Al$_x$Ga$_{1-x}$N on GaN above the critical thickness and can already be observed in a sample twice thinner and grown under the same conditions (not shown). In contrast, in the presence of Mn (sample \#F), the domes already seen in GaN:Mn are detected, but there is no evidence of cracks in the field of view, as shown in fig.~\ref{fig:fig1}(f). Optical microscopy in reflexion mode also reveals the presence of cracks in the Mn-free sample, while a whole 2" wafer grown in the presence of Mn is completely crack-free, pointing to an Al$_x$Ga$_{1-x}$N layer perfectly strained with the GaN buffer.

\begin{figure}[ht]
  \includegraphics[width=.9\linewidth]{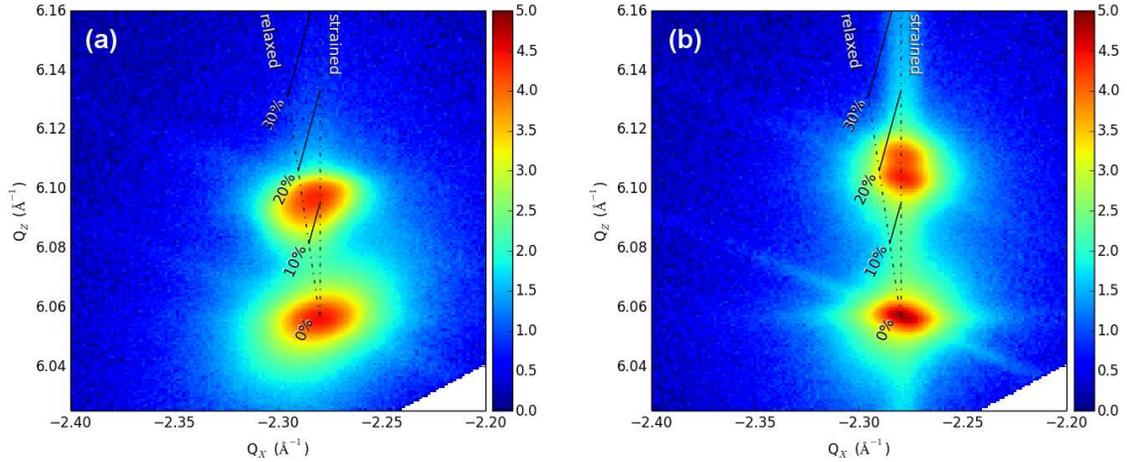}
  \caption{Reciprocal space maps around GaN and Al$_x$Ga$_{1-x}$N$(\overline{1}015)$ for: (a) sample \#E and (b) sample \#F. The intensity is reported in logarithmic scale. A vertical dashed line along the GaN $(\overline{1}01l)$, and an oblique dashed line joining experimental GaN and AlN $(\overline{1}015)$ are drawn as guides to the eye. The isoconcentration lines are indicated as continuous lines between the strained and relaxed states.}
  \label{fig:fig2}
\end{figure}

This result is confirmed by x-ray diffraction experiments. Reciprocal space maps have been measured around the $(\overline{1}015)$ reflexion of GaN and Al$_x$Ga$_{1-x}$N and are reported in fig.~\ref{fig:fig2} for samples \#E and \#F. The shape and position of this peak appear to be different for the two samples: the center of the peak is shifted towards lower in-plane lattice parameters in the case of the Mn-free sample.  Here the average lattice parameter of the layer does not fit the one of GaN, indicating a plastic relaxation of the crystal lattice. In fact the peak is neither aligned with the dashed line which corresponds to a fully relaxed layer nor with the one corresponding to a fully strained state, pointing to an intermediate strain state. Furthermore, the peak is particularly broad, actually spreading over the whole range between strained and relaxed state. In comparison, in the presence of Mn, the $(\overline{1}015)$ reflexion of AlGaN is very narrow in $Q_x$, and vertically aligned with the $(\overline{1}015)$ of GaN, confirming the strained state of the layer already evidenced by surface microscopy. Despite the Al concentration is comparable in the two samples, the peak of the layer containing Mn and not relaxed is shifted towards higher values of $Q_z$ due to the limited compressibility of the material. In addition, this peak exhibits an unexpected broadening along the $Q_z$ direction, suggesting the presence of Al$_x$Ga$_{1-x}$N with different Al concentrations. In order to quantify the Al content in the films from the position of the $(\overline{1}015)$ peak, we assume a linear variation of the out-of-plane lattice parameter with the Al content in the whole range of concentrations from GaN to AlN (Vegard's law). In the completely relaxed case, the Al concentration is given by $x_\mathrm{Al}=\frac{c_\mathrm{AlGaN}-c_\mathrm{GaN}}{c_\mathrm{AlN}-c_\mathrm{GaN}}$. In the perfectly strained case, it is necessary to add a prefactor to take into account the elongation of the lattice along the $c$ direction when the crystal is elongated along the $a$-axis. The Al concentration is then obtained through $x_\mathrm{Al}=\frac{1-\nu}{1+\nu}\frac{c_\mathrm{AlGaN}-c_\mathrm{GaN}}{c_\mathrm{AlN}-c_\mathrm{GaN}}$ where $\nu$ is the Poisson coefficient (0.19 and 0.21 for GaN and AlN respectively). The calculated Al concentration is thus in the considered sample $(12\pm1)$\% for the partially relaxed Mn-free layer (sample \#E). For the Mn-containing layer (sample \#F), the two main peaks related to Al$_x$Ga$_{1-x}$N correspond to Al contents of 12.8\% and 14.3\%, respectively. 

\begin{figure}%
\includegraphics[width=.7\columnwidth]{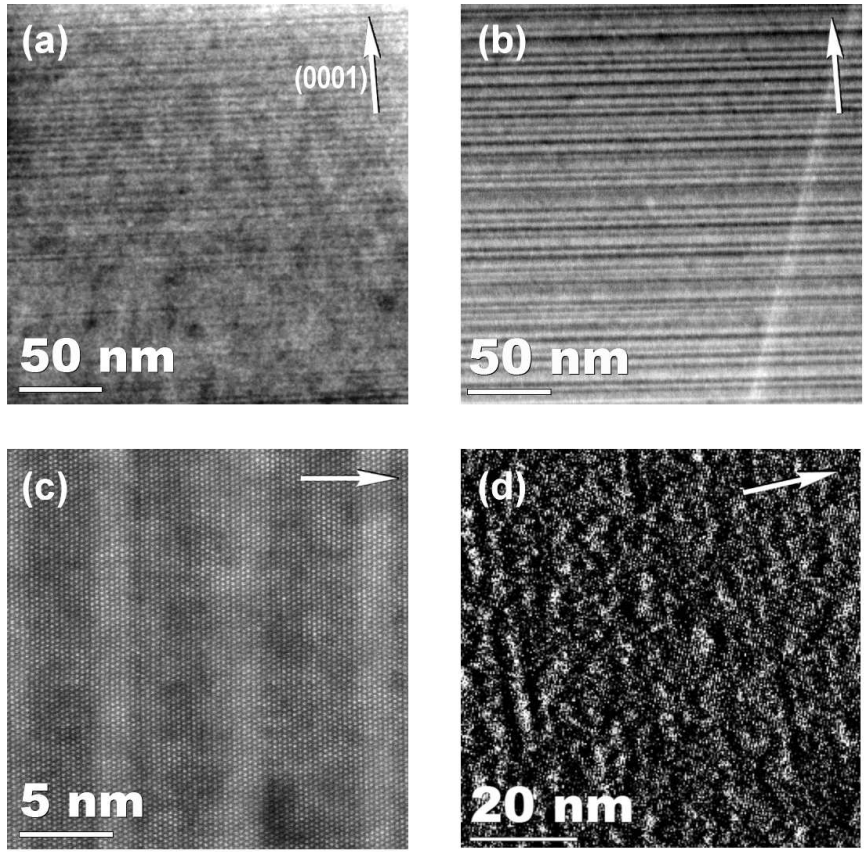}%
\caption{TEM of the Al$_x$Ga$_{1-x}$N:Mn layer of sample \#F in cross-section measured along the [11$\overline{2}$0] zone axis: (a) low magnification conventional TEM; (b) low magnification HAADF; (c) high resolution HAADF; (d) Al chemical map from energy filtered TEM measured at Al $K$-edge; in all panels, the (0001) direction is indicated by an arrow.} %
\label{fig:fig3}%
\end{figure}

In order to shed light on the origin of the different Al concentrations detected by XRD in the films doped with Mn, the nature of the Al inhomogeneity and the role of Mn on the Al segregation, the layers have been investigated with (HR)TEM. In fig.~\ref{fig:fig3}(a) a low magnification transmission electron micrograph of the Al$_{0.12}$Ga$_{0.88}$N:Mn layer reveals  along the \emph{c}-axis the presence of a quasi-periodic structure, which is not observed in the Mn-free samples. An analysis over 40 sub-layers gives an average thickness of 2.9\,nm with a standard deviation of 0.73\,nm. No significant difference in the thickness of darker and brighter layers could be found. Both in TEM and in XRD, the non-perfect periodicity of the superlattice does not allow to resolve the satellite peaks characteristic of a superlattice. To discriminate between the contrast due to diffraction effects from the one induced by composition contrast, the same layer has been measured in HAADF in STEM mode. Within this imaging technique, electron diffusion measured at a high angle is decisively dependent on the mass of the diffusing elements. The contrast observed can therefore be directly correlated with a mass contrast, the Al-rich areas being darker than the Ga-rich ones. The HAADF measurements are reported in fig.~\ref{fig:fig3}(b) and (c) and show a similar patterning as in fig.~\ref{fig:fig3}(a). This suggests that the contrast observed in conventional TEM is induced by a modulation in the Al concentration. This result is also confirmed by the energy filtered image displayed in fig.~\ref{fig:fig3}(d). This image has been acquired in EFTEM, around the Al $K$-edge and the bright areas correspond to Al-rich regions. This element-specific technique confirms that the mass contrast observed in HAADF originates indeed from Al segregation, and not only from a variation of the density of the material, which could be eventually induced by the presence of defects. 

A similar self-structuration of ternary alloys in superlattice-like heterostructures was already reported for the growth of Al$_x$Ga$_{1-x}$N on GaN\cite{Korakakis1997, Doppalapudi1998} and AlN \cite{Iliopoulos2001,Gao2006,Wang2006,Kim2010}. Particularly remarkable here, is not the segregation of Al itself, but rather the fact that we are able to trigger the segregation and the self-assembling of layered structures through the presence of Mn. In order to figure out the underlying mechanism, one should consider that Mn is here playing the role of an efficient surfactant, mostly accumulating at the growing surface. Therefore, Mn behaves like the cations Ga and Al and affects radically the equilibrium between the metal (Al or Ga) and nitrogen, $i.e.$ the III/V ratio, known to have a key role in the decomposition into superlattices\cite{Iliopoulos2001,Wang2006}. 

\begin{figure}%
\includegraphics[width=.9\columnwidth]{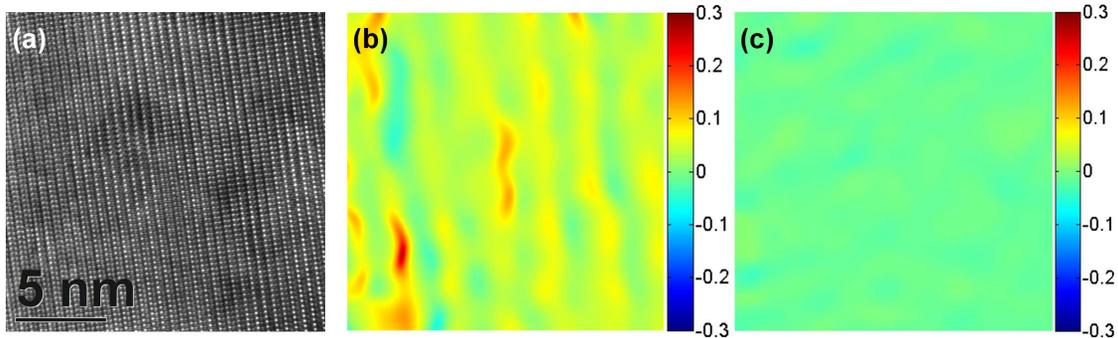}%
\caption{Strain analysis for Al$_{0.12}$Ga$_{0.88}$N:Mn (sample \#F) obtained through the geometrical phase analysis method\cite{Hytch1998}: (a) high resolution transmission electron micrograph in cross section taken along [11$\overline{2}$0] zone axis; (b) and (c) corresponding strain maps along the \emph{c}- and \emph{a}-direction respectively.} %
\label{fig:fig4}%
\end{figure}

Since the layer (sample \#F) is pseudomorphically grown on GaN, the modulation of the Al concentration is expected to be accompanied with a modulation of the \emph{c}-parameter.
In order to establish the strain distribution in the layer, a geometrical phase analysis (GPA) according to the technique developed by H\"ytch \emph{et al.}\cite{Hytch1998} has been performed. The GPA is implemented from the high resolution micrograph measured in the [11$\overline{2}$0] zone axis and reported in fig.~\ref{fig:fig4}(a). Since the phase modulation takes place in the whole area of imaging, and since the strain state in this area is \emph{a priori} unknown, the average phase has to be taken as a reference for the calculation of the strain. The strain maps are calculated along the \emph{c}- and  \emph{a}-axis and represented in fig.~\ref{fig:fig4}(b) and (c) respectively. 

The absence of strain contrast along the \emph{a}-direction is coherent with the pseudomorphic character of the layer already evidenced by XRD. The quasi-periodic structure detectable in the strain map calculated along the \emph{c}-direction indicates that the \emph{c} lattice parameter is modulated, confirming the periodic variation of Al content already observed in HAADF. The average lattice parameter measured in the HRTEM considered in fig.~\ref{fig:fig4}(a) is 5.128\,\AA\, which -- taking into account the strained character of the layer -- corresponds to an Al concentration of 13.1\% . The average negative strain (compressive) and positive strain (tensile) in fig.~\ref{fig:fig4}(b) are -1.94\% and
2.84\%, and correspond to a \emph{c}-parameter of 5.029\,\AA ~ and 5.274\,\AA, respectively. Considering that the layer is pseudomorphically grown on GaN, a \emph{c}-parameter of 5.029\,\AA\,gives an Al concentration of about 35.0\% which is far above the one extracted from XRD, while 5.274\,\AA\,is greater than the \emph{c}-parameter of relaxed GaN (5.185\,\AA). This unexpectedly large lattice parameter can be explained by the presence of local defects locally distorting the lattice. On the other hand, it has to be considered that the non-perfect periodicity of the layered structure is likely to generate a broad diffraction line containing Al$_x$Ga$_{1-x}$N and GaN peaks, rather than well defined satellites, as evidenced in fig.~\ref{fig:fig2}(b). A similar weak effect can be also appreciated in the fast Fourier transform (FFT) of some of the HRTEM pictures, particularly in two beam condition, where the (0002) peak gets elongated in the (000l) direction. Such a pattern in the FFT of fig.~\ref{fig:fig4}(a), and particularly  the fact that a part of it may get lost in selecting the diffracted spot used for the strain calculation,  may produce a systematic error in the quantification of the strain, and consequently of the lattice parameter. This error could give too high (resp. low) lattice parameter for low (resp. high) Al content layers in the superlattice. 

\section{Conclusions}

We have shown that Mn acts as a surfactant in the MOVPE of the technologically strategic compounds GaN and Al$_x$Ga$_{1-x}$N, affecting the surface morphology and the crystalline arrangement. Most remarkably, in the case of low Al concentrations we have found that Mn induces the segregation of Al in the Al$_x$Ga$_{1-x}$N:Mn films and promotes the self-assembling of layered superlattice-like structures. In a larger perspective and particularly in the case of reactive crystal growth techniques -- among which MOVPE -- the use of appropriate surfactants, and their potential of affecting the balance between the different species involved in the growth process, can open a new route to control the segregation of selected elements in ternary and more complex alloys and to promote the self-assembling of functional heterostructures.

\begin{acknowledgement}
The work was supported by the European Research Council (ERC Advanced Grant 227690), by the Austrian Science Fundation (FWF Projects 22477, 24471) and by WRC EIT+ within the project NanoMat (P2IG.01.01.02-02-002/08) co-financed by the European Operational Programme Innovative Economy (1.1.2).
\end{acknowledgement}




\bibliography{AlGaNMn-THD-v1.4}

\section{Table of content graphics}
\begin{figure}%
\includegraphics[width=89mm]{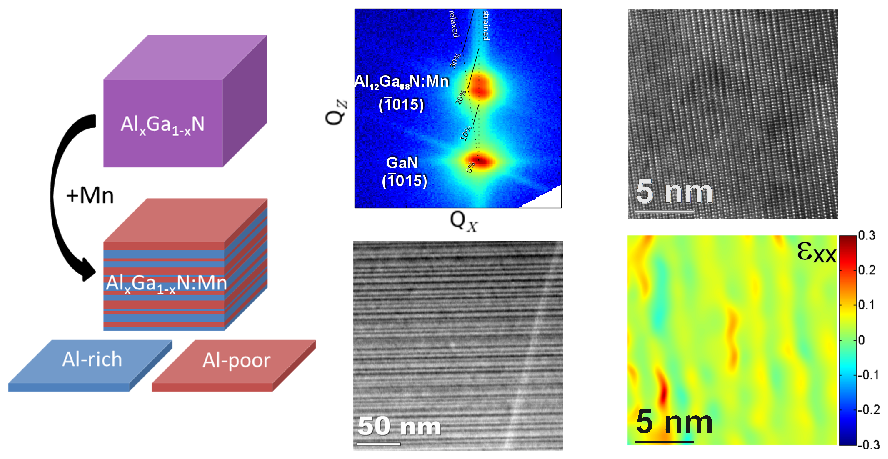}%
\caption{} %
\label{fig:TOC}%
\end{figure}

\end{document}